\documentclass[unnumbered]{trbunofficial}

\usepackage{booktabs}
\usepackage{graphicx}
\usepackage{tikz}
\usetikzlibrary{arrows.meta, positioning, shapes.geometric}
\usepackage{placeins}
\graphicspath{{figures/}}

\begin{document}


\title{Treating Statewide CORS Networks as Spatially Distributed Sensors for GNSS Integrity Monitoring under Unintentional and Deliberate Threats}

\TRBauthor*{Minhaj Uddin Ahmad}{Department of Civil, Construction and Environmental Engineering, The University of Alabama}{mahmad12@crimson.ua.edu}[Tuscaloosa, AL, 35487][0000-0002-8600-2685]

\TRBauthor{Sagar Dasgupta}{Department of Civil, Construction and Environmental Engineering, The University of Alabama}{sdasgupta@ua.edu}[Tuscaloosa, AL, 35487][0000-0001-8491-662X]

\TRBauthor{Muhammad Sami Irfan}{Department of Civil, Construction and Environmental Engineering, The University of Alabama}{mirfan@crimson.ua.edu}[Tuscaloosa, AL, 35487][0000-0002-1116-935X]

\TRBauthor{Mizanur Rahman}{Department of Civil, Construction and Environmental Engineering, The University of Alabama}{mizan.rahman@ua.edu}[Tuscaloosa, AL, 35487][0000-0003-1128-753X]

\TRBauthor{Mashrur Chowdhury}{Glenn Department of Civil Engineering, Clemson University}{mac@clemson.edu}[Clemson, SC, 29631][0000-0002-3275-6983]

\TRBauthor{Thejesh N. Bandi}{Department of Physics \& Astronomy, The University of Alabama}{tbandi@ua.edu}[Tuscaloosa, AL, 35487][0000-0001-7463-7301]

\AuthorHeaders{Ahmad, Dasgupta, Irfan, Rahman, Chowdhury, and Bandi}


\maketitle

\section{Abstract}

\textbf{Objectives:} State departments of transportation (DOTs) in the United States increasingly rely on statewide continuously operating reference station (CORS) networks to support high-precision Global Navigation Satellite System (GNSS)-based positioning and timing, which are essential to intelligent transportation system applications. Although these networks are operated as positioning infrastructure, they also provide continuous observations that can support regional GNSS integrity monitoring. This study develops and demonstrates a framework that treats a statewide CORS network as a spatially distributed sensor system for identifying unintentional (i.e., environmental vulnerabilities) and intentional (i.e., deliberate cyber threats) interference events where GNSS measurements deviate from expected patterns.

\hfill\break
\noindent\textbf{Methods:} We develop a graph-based Network Consistency Framework (NCF) that evaluates each station against its spatial neighborhood using four complementary consistency metrics: neighborhood residual, spatial gradient, residual, and graph smoothness. The metrics are combined into a Network Consistency Index (NCI). The framework is demonstrated using two consecutive days of four-constellation observations from 50 stations of Alabama DOT-maintained CORS network, using changes in vertical total electron content ($\Delta$VTEC) and the Rate of TEC Index (ROTI) as spatially coherent observables.

\hfill\break
\noindent\textbf{Findings:} The framework successfully quantified network-wide spatial consistency and identified localized anomalies through the Network Consistency Index (NCI). The detected anomalies represent stations whose observations deviated from the surrounding regional network, indicating potential integrity issues. Determining whether these anomalies originate from receiver faults, localized interference, spoofing, or other causes requires further investigation.

\hfill\break
\noindent\textbf{Novelty:} This study introduces the concept of treating statewide CORS networks as regional GNSS integrity observatories and presents the Network Consistency Framework and Network Consistency Index for graph-based spatial integrity monitoring.

\hfill\break
\noindent\textbf{Practical Applications:} Transportation agencies can implement the presented framework using existing CORS observations without additional hardware to continuously monitor network integrity, identify localized interference and receiver faults, and provide a regional cross-check to support GNSS spoofing detection.

\newpage

\section{Introduction}

Many aspects of transportation systems increasingly rely on precise Global Navigation Satellite System (GNSS) positioning. State departments of transportation (DOTs) and their partners rely on centimeter-level real-time kinematic (RTK) and post-processed positioning for highway and boundary surveying, construction staking, automated machine guidance (AMG) on grading and paving equipment, mobile and aerial mapping, asset inventory, the construction of digital twins, and unmanned aerial vehicle (UAV) based surveys. A recent Caltrans preliminary investigation surveying state practice found that DOTs use precise positioning and timing across a broad range of applications ``from navigation and infrastructure development to land surveying,'' and identified automated transportation as a leading motivation for expanding statewide real-time GNSS networks \citep{Caltrans2025}.

To provide these services, state agencies operate statewide GNSS real-time networks (RTNs) consisting of continuously operating reference stations (CORS). Each reference station is essentially a continuously operating GNSS receiver whose antenna is installed at a precisely surveyed reference point. The Caltrans survey reports that in ten of thirteen responding agencies, the DOT owns or co-owns and maintains the reference stations, at system costs of roughly \$1--10\, million and annual budgets near \$1\, million \citep{Caltrans2025}, which is a major public investment. These networks are primarily viewed and instrumented as precise positioning infrastructures. Additionally, researchers across multiple disciplines, such as geophysics and meteorology, use these networks to derive various products of interest~\citep{snay2008continuously}. 

GNSS signals arrive at the Earth's surface with extremely low signal strength, and civil signals are unauthenticated, except for Navigation Message Authentication (NMA) in the new Galileo constellation \cite{nicola2022galileo}. The power level at Earth's surface is on the order of $10^{-16}$ watts, which is easily overpowered by an adversary. An inexpensive transmitter can broadcast a counterfeit constellation that steers a receiver's reported position or time to attacker-chosen values \citep{humphreys2008assessing}. This is a cybersecurity problem, and a particularly growing one. The EU STRIKE3 project cataloged hundreds of thousands of reported interference incidents, more than 10\% of which were deliberate \citep{thombre2018gnss}. Spoofing and jamming have been documented at a national scale \citep{C4ADS2019}. Researchers have found a relationship between geopolitical instability and GNSS spoofing and jamming \citep{pik2025predicting}, indicating the need for resilient GNSS infrastructure and spoofing awareness. For transportation applications, a spoofed signal that feeds an automated vehicle, a survey of record, or a timing-dependent system constitutes a failure of safety, integrity, and trust.

Numerous studies have addressed GNSS integrity by evaluating each receiver in isolation. Receiver-level methods test a single receiver's own measurements for self-consistency \cite{blanch2015baseline}. Interference monitors deployed on national reference networks compare each station's signal-to-noise ratio against a historical model of that same station \cite{abraha2024gnss}. While both classes of method are effective within their scope, neither exploits a property that reference networks possess by construction. The stations observe a shared geographical environment and should therefore agree with one another on spatially coherent quantities, such as ionospheric parameters. Multi-receiver spoofing detection does use agreement between separated receivers as its statistic, but it has been developed for short baselines, on the order of tens of meters to roughly one kilometer, over which atmospheric delays are common to all receivers and are deliberately differenced away \cite{khanafseh2017ephemeris}. As a result, there remains a gap in the literature concerning the intermediate regime, where permanently surveyed stations separated by tens of kilometers are treated as a single distributed sensor over a state-sized area.

The objective of this study is to use the statewide CORS infrastructure that DOTs already own and operate as a continuously running regional GNSS integrity sensor for transportation users, and, in particular, as a spatial cross-check that indicates potential GNSS spoofing. The premise of the approach is that GNSS-derived products, such as ionospheric parameters, evolve as spatially coherent fields over large geographical areas. This coherence is the same assumption that makes network RTK work \citep{rizos2003reference}. A counterfeit signal from a terrestrial spoofer or local radio-frequency interference (RFI), by contrast, acts at a single site. The affected station consequently ceases to match the field defined by its neighboring stations, while those neighbors continue to agree among themselves. Rather than testing whether a station appears anomalous relative to its own history, we test whether it appears anomalous relative to its spatial neighborhood at the same epoch, thereby converting the network from a collection of independent sensors into a single distributed sensor with a built-in, network-wide baseline. Hence, the baseline is not defined by historical data; rather, it is defined by the instantaneous measurements of surrounding stations. 

The contributions of this paper are as follows. 

\begin{itemize}
    \item Introduce the concept of treating statewide CORS networks as spatially distributed regional GNSS integrity sensors for civil and transportation use.
    \item Present a Network Consistency Framework (NCF), a method that represents the RTN as a graph and quantifies each station's spatial agreement with its neighborhood.
    \item Define a family of four complementary spatial consistency metrics: (1) neighborhood residual, (2) local spatial gradient, (3) prediction residual, and (4) graph-smoothness.
    \item Introduce a Network Consistency Index (NCI), a single composite indicator that summarizes spatial consistency across multiple metrics.
    \item Demonstrate the framework under nominal operating conditions on the Alabama DOT CORS network.
\end{itemize}
 
The remainder of this paper is structured as follows. The next section reviews the related literature on receiver-level and network-level GNSS integrity monitoring. We then develop the framework, NCF, and the composite index, NCI. Subsequently, we present the dataset, the data availability accounting, and report results for two full days of four-constellation observations. Finally, we conclude with a summary of findings and a set of practical implications for transportation agencies.

\section{Literature Review}
 
We express spoofing detection and integrity monitoring strategies into four categories based on their operational context and capability: 1) receiver-level integrity monitoring and spoofing detection; 2) network-based augmentation systems and their fault screening; 3) geodetic reference networks operated as interference sensors; and 4) multi-receiver and cooperative spoofing detection.
 
\subsection{Receiver-Level Integrity Monitoring}
The established approach to GNSS integrity is for a receiver to test its own measurements for internal consistency. Receiver Autonomous Integrity Monitoring (RAIM) and its multi-constellation successor, Advanced RAIM, use measurement redundancy to detect and exclude faulty ranges and to bound the resulting position error \citep{parkinson1988autonomous,blanch2015baseline}. A substantial body of work has extended this idea to interference and spoofing by monitoring quantities that a counterfeit signal is likely to disturb. Representative approaches include automatic gain control as a direct indicator of received power anomalies \citep{akos2012s}, carrier-to-noise density and correlation function shape as real-time signal quality metrics \citep{borio2015real}, and comparison against a non-GNSS reference, such as an inertial measurement unit, in order to detect a spoofer capable of tracking vehicle motion \citep{tanil2017ins}. Spatial processing at the receiver has also proven effective. For instance, \citep{broumandan2016overview} surveyed multi-antenna methods that estimate the direction of arrival of each tracked signal, and showed that signals originating from a single terrestrial transmitter can be discriminated on that basis, and Rothmaier et al.\ developed a spatial processing formulation for the same purpose \citep{rothmaier2021gnss}. General surveys of the threat and of these countermeasures are available \citep{psiaki2016gnss}.
 
While these methods have been shown to work well, the available evidence for testing is limited to a single receiver. These single-receiver detection strategies generally work well when the receiver is first attacked, but a receiver already fully captured by a self-consistent spoofer has little reason to suspect. Some receiver-level defenses, such as multi-element antenna arrays, require additional hardware.

\subsection{Network-Based Augmentation and Fault Screening}
 
System-level integrity monitoring uses multiple reference receivers rather than a single receiver. Ground-Based Augmentation Systems (GBAS) operate redundant reference receivers at an airport, detect navigation-related faults, compute error bounds, and communicate them to users, and such architectures are maintained by the Federal Aviation Administration (FAA) in the United States \citep{faa_satnav, khanafseh2017ephemeris}. Satellite-Based Augmentation Systems (SBAS) and Precise Point Positioning (PPP) services estimate a state space representation of GNSS error sources from a network of reference stations, and in doing so they already perform cross-station consistency checks, in that an observation whose residual in the network least-squares solution exceeds a screening threshold is removed so that the anomaly does not propagate into the broadcast corrections \citep{wubbena2005ppp,sc1996minimum}.
 
While these systems demonstrate that multi-receiver fault detection is both feasible and operationally proven, three limitations remain. First, GBAS operates within a local service volume near an airport rather than over a large geographic area. Second, the screening performed by SBAS and PPP services is intended to protect the correction product, so the anomalous station is discarded, and the event that produced the anomalous measurement residual is neither localized nor characterized. Third, the large networks supporting SBAS and PPP services are continental or global in extent and sparsely distributed, and they therefore lack the spatial resolution needed to attribute an anomaly to a specific location.
 
\subsection{Reference Networks as Interference Sensors}
Several national reference networks have been instrumented as interference sensors. The Swedish SWEPOS and Finnish FinnRef networks operate automated systems that monitor each station's signal-to-noise ratio against a historical model and issue an alert when an anomalous event is detected, which has proven to be an effective means of identifying jamming and radio-frequency interference \citep{abraha2024gnss, nikolskiy2020gnss}. Subsequent studies have applied machine learning techniques to the same per-station quality monitoring problem \citep{lebrun2021gnss}. Quality-control software such as now-deprecated TECQ and new G-Nut/Anubis, widely used on reference networks, reports per-station statistics in a similar spirit and does not perform cross-station spatial inference \cite{vaclavovic2015g}.
 
Studies that utilize the spatial properties have generally had the purpose of characterizing the ionosphere rather than monitoring integrity. For instance, \citep{sokolova2023high} combined a cluster of CORS receivers with shorter-baseline supplemental stations to isolate and analyze ionospheric spatial gradients and to quantify the degree of spatial decorrelation between stations during periods of increased ionospheric activity. Their study used the network geometry as a measurement instrument for a geophysical quantity, rather than as a consistency test.
 
While the mentioned network monitoring systems have demonstrated real operational value, they remain, in essence, banks of single-station detectors. Each station is compared only with its own past, so a fault that is unremarkable relative to that station's history is not flagged.
 
\subsection{Multi-Receiver and Cooperative Spoofing Detection}

A separate line of work uses agreement between physically separated receivers as the detection statistic, and it is the body of literature most closely related to the present study. \citet{tippenhauer2011requirements} established the geometric requirements that an attacker must satisfy to spoof several victims simultaneously, and showed that while any number of receivers can be driven to a single common location, preserving the victims' relative formation restricts the attacker to very few transmission positions. Detection methods built on this asymmetry compare the reported positions or the differential ranges of receivers whose relative geometry is known. Swaszek et al.\ analyzed a family of detectors that monitor whether the position solutions of two or more receivers at known relative positions abnormally coincide and reported their detection performance under realistic error assumptions \citep{swaszek2013analysis,swaszek2013spoof}. A widely adopted formulation applies a generalized likelihood ratio test to double-differenced pseudorange or carrier-phase measurements collected from multiple independent receivers \citep{jahromi2016gnss,wen2019spoofing, stenberg2022results}. More recently, \citet{chen2025gnss} replaced the likelihood ratio test with a statistic based on the spatial distribution of double-differenced pseudoranges and demonstrated improved discrimination of partially spoofed scenes, along with greater tolerance to multipath. At larger spatial scales, crowdsourced approaches have used opportunistic consumer receivers or aircraft surveillance broadcasts to detect and localize interference sources over wide areas \citep{jansen2018crowd,strizic2018crowdsourcing}.
 
While these studies share our premise that separated receivers can be checked against one another, they operate in a regime that differs from ours in a specific and consequential way. The receiver separations range from tens of meters to roughly one kilometer, and the double-differencing step that makes them tractable removes the satellite clock bias and the ionospheric and tropospheric delays, on the grounds that these terms are common to all the receivers on short baselines. The quantity canceled by the previous studies is precisely the one our study investigates. At the 21--55\, km inter-station spacing of a statewide RTN, atmospheric delay does not cancel between neighboring stations; it instead varies smoothly, and that smooth variation constitutes the field against which each station is tested. The two approaches are therefore complementary rather than competing. Short-baseline methods test whether a set of receivers is mutually consistent with a common counterfeit transmission point, whereas the framework presented here tests whether a station is consistent with the geophysical field defined by its neighbors.
 
The spatial scale of implementation is relevant to integrity monitoring. The radius within which a terrestrial spoofer can capture a receiver is short. For an omnidirectional transmitter at 0\,dBm, published link-budget analysis places the radius within which a receiver is reliably overwhelmed at roughly 120\,m, and the radius within which capture becomes possible at a few kilometers \citep{chen2025gnss}. Both figures are well below the median inter-station spacing of the network studied here, so a single terrestrial spoofer cannot plausibly capture two stations simultaneously.
 
Finally, ionospheric observables have themselves been proposed for spoofing detection, although so far at the single-receiver level. A dual-frequency receiver can estimate total electron content along each signal path and compare it against an external reference, such as a global ionosphere map from online sources, on the reasoning that a simulator-based spoofer lacking access to real-time ionospheric information cannot reproduce the authentic delay in a timely manner \citep{wu2025ionospheric}. That study supports the choice of ionospheric quantities as spoofing-sensitive observables. It differs from our approach in the reference against which the comparison is made. It compares one receiver against a global model from other sources, whereas we compare stations against one another at regional resolution and require no external correction product.
 
In summary, receiver-level monitors are confined to the evidence that one receiver can gather; network-based augmentation systems use several receivers, but either over a local service area or in order to protect a differential correction product, and they discard rather than localize anomalies; reference networks operated as interference sensors compare each station only with its own history; and multi-receiver spoofing detection tests agreement between separated receivers, but in a short-baseline distances that cancels the atmospheric effects. Consequently, the concept of treating a statewide network of permanently surveyed stations, spaced tens of kilometers apart, as a single distributed sensor, in which neighboring stations define one another's expected behavior and deviations from that are localized and scored, remains underexplored. This gap motivates the framework developed in the remainder of this paper.
 
\section{The Network Consistency Framework}
 
This section develops the framework to evaluate network consistency. We first state the problem formulation and the spatial representation, then specify which GNSS observables are admissible, and finally define four complementary consistency metrics and the composite index that summarizes them.

\begin{figure}[htbp]
  \centering
  \includegraphics[width=0.80\linewidth]{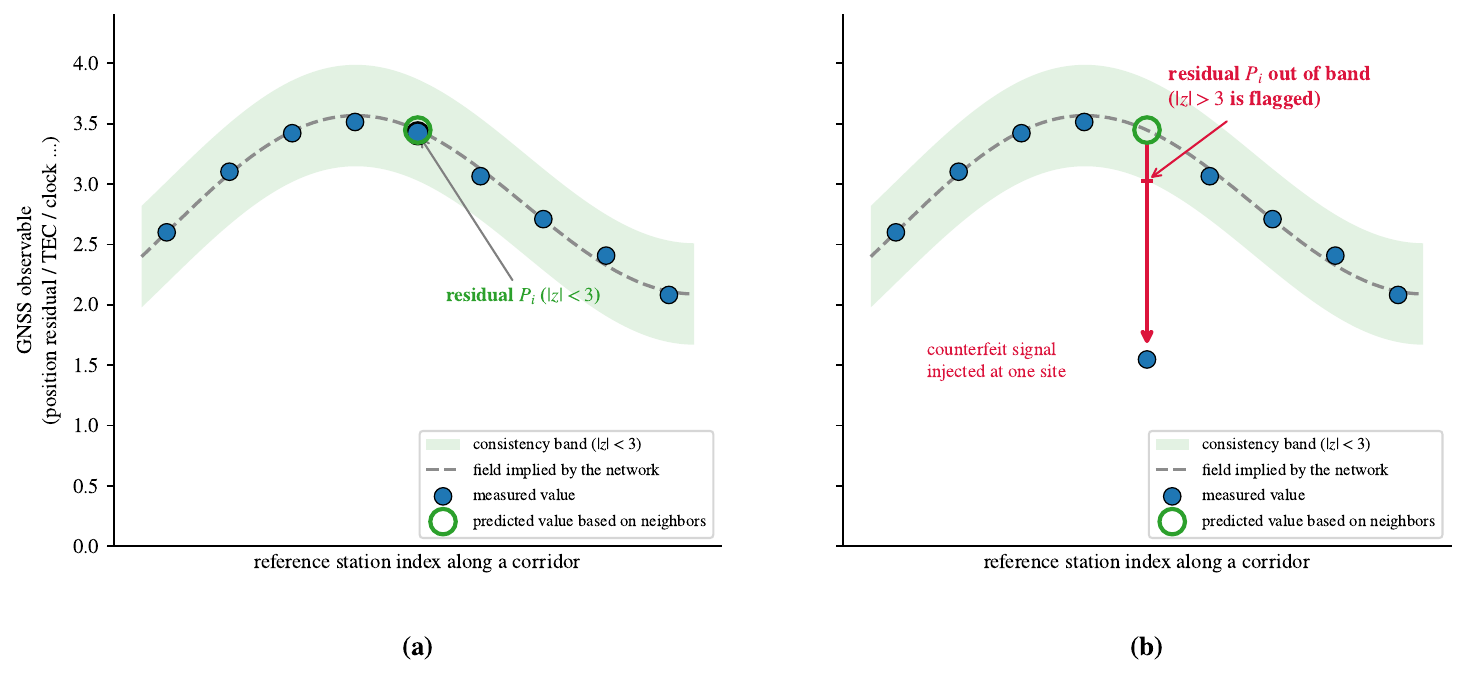}
  \caption{Principle of the network cross-check. (a) Under nominal conditions (b), a single-site spoof breaks that coherence.}
  \label{fig:concept}
\end{figure}

Figure~\ref{fig:concept} illustrates the conceptual assumption based on which the framework is developed, following the literature review. An adversary who wished to remain hidden within the network would have to satisfy three coupled constraints at every affected station simultaneously: 1) reproducing the exact geometric range and range rate to each satellite for that station's surveyed coordinates and clock; 2) injecting the correct, position-specific atmospheric delays, so that the reconstructed parameters match the smooth surface that the neighbors define; and 3) maintaining mutual consistency between code and carrier phase on every frequency as the true satellites move across the sky. Consequently, evasion would require coordinated, mutually consistent spoofing across the entire region. As discussed earlier, the capture radius of a plausible terrestrial spoofer is well below the median inter-station spacing of this network, so the coordinated case requires a corresponding number of separate transmitters rather than a single more powerful one.
 
\subsection{Problem Formulation}
 
Let $N$ reference stations observe a set of satellite constellations. For a chosen observable, let $f_i \in \mathbb{R}$ denote the scalar value reported at station $i$ over an analysis window, and let $\mathbf{p}_i \in \mathbb{R}^2$ denote its projected map coordinates. For each station in turn, the framework asks how far $f_i$ departs from the value that the remaining stations would be expected to yield at $\mathbf{p}_i$. A station in agreement with its neighborhood scores near zero, and a station that departs from its neighborhood scores a higher value.
 
\subsection{Spatial Representation}
 
We represent the network as an undirected graph $\mathcal{G}=(V,E)$ whose nodes $V$ are the stations and whose edges $E$ encode spatial neighbor relationships, taken here to be the $k$ nearest neighbors of each station in the map projection. Each node carries the observable value $f_i$, and each edge carries a weight $w_{ij}$ that decreases with the inter-station distance $d_{ij}=\lVert\mathbf p_i-\mathbf p_j\rVert$. This construction follows the standard setting in signal processing on graphs \citep{Shuman2013}, and it allows us to borrow well-defined notions of smoothness and residual analysis from that literature.
 
We set $k=6$ throughout, which is the smallest neighborhood that keeps the graph connected for this station geometry while remaining local enough that a single anomalous neighbor cannot dominate the median in Equation~\eqref{eq:resid}. 
 
\subsection{Candidate Observables}
 
Not every station-derived measurement is admissible as input to a spatial consistency test, and the distinction is important enough to state before any results are presented. We therefore divide the observables into two classes.
 
\emph{Class A observables} are geographically interpolable. They are governed by the atmospheric state over the region rather than by the hardware at any one site, and consequently form fields that either vary smoothly across the network or remain approximately uniform under nominal conditions. For such observables, spatial interpolation from neighboring stations is expected to be close to the measured value.
 
\emph{Class B observables} are site-specific and interference-sensitive. The mean $C/N_0$ level used here is dominated by station-specific hardware and the radio environment within a short distance of the antenna. Neighboring stations carry no information about the nominal value of this quantity at a given site, so interpolating it between stations is not meaningful. Other quantities governed by the antenna's immediate environment, such as code multipath and single-point positioning scatter, fall into the same class by the same reasoning. Class B observables are nevertheless retained, for two reasons. First, a depression in $C/N_0$ relative to the same station's own recent behavior is the principal means through which jamming and broadband interference become visible, and no Class A observable carries that information. Second, Class B observables provide evidence external to the spatial test, which can be used for further investigation.

\subsubsection{Total electron content in the ionosphere}
Total electron content (TEC) is a Class A observable used in this study. Free electrons in the ionosphere cause dispersion and code delay in GNSS signals. First-order ionospheric delay on a code measured at frequency $f$ is $I_f = (40.308/f^2)\,\mathrm{STEC}$, where the slant total electron content (STEC) is expressed in electrons\,m$^{-2}$ and 1~TECU $=10^{16}$~el\,m$^{-2}$ \citep{parkinson1996global}. As this term is dispersive in nature, differencing the two code pseudoranges, which is known as the geometry-free combination, cancels the geometry, the clock, and the troposphere, and isolates the ionospheric contribution:
\begin{equation}
\rho_2 - \rho_1 = 40.308\,\mathrm{STEC}\left(\frac{1}{f_2^{2}}-\frac{1}{f_1^{2}}\right) + c\,(b_r + b^s),
\end{equation}
so that
\begin{equation}
\mathrm{STEC} = \frac{1}{40.308}\,\frac{f_1^{2}f_2^{2}}{f_1^{2}-f_2^{2}}\,\bigl[(\rho_1 - \rho_2) - c\,(b_r + b^s)\bigr],
\end{equation}
where $b_r$ and $b^s$ are the receiver and satellite differential code biases (DCBs). A DCB is the difference in hardware signal delay between the two frequencies, arising in the receiver's analog front end and in the satellite's transmitter. It is unknown; it takes different values at each receiver and on each satellite, and it is stable over hours to days. The geometry-free carrier combination $\lambda_1\varphi_1-\lambda_2\varphi_2$ measures the same quantity with far lower noise but carries an unknown ambiguity, and smoothing the code measurement with carrier measurement over each continuous tracking arc yields a low-noise, unambiguous STEC \citep{ciraolo2007calibration}. A thin-shell ionospheric model at height $H$ converts the slant quantity to vertical total electron content (VTEC),
\begin{equation}
\mathrm{VTEC} = \mathrm{STEC}\,\cos z', \qquad \sin z' = \frac{R_E}{R_E+H}\cos(\mathrm{el}),
\end{equation}
in which $\mathrm{el}$ is the elevation angle and $R_E$ is the Earth radius; we use $H=350$\,km and a $25^\circ$ elevation mask.
 
Absolute VTEC therefore requires that the DCBs be known. However, if we take the time difference in VTEC, we get $\Delta$VTEC. This difference cancels $b^s$, and it cancels $b_r$ as well whenever the receiver bias is stable between the time intervals, leaving only the ionospheric change. A receiver whose DCB jumps between days violates that assumption.
 
\subsubsection{Rate of change of total electron content}
The rate-of-TEC index (ROTI) is the standard deviation of $\mathrm{d\, STEC}/\mathrm{d}t$ over the analysis window. It is worth noting that ROTI is directly related to geomagnetic storms \citep{pi1997monitoring}. It is also a Class A observable. Under geomagnetically quiet conditions, ROTI is small and close to uniform across a region.
 
\subsection{Spatial Consistency Metrics}
 
For a given Class A observable field $\{f_i\}$ we define four metrics per station. They are complementary by construction. The first and third measure whether station $i$ itself departs from its neighborhood, the second measures whether $i$ lies in a region of abrupt spatial change, and the fourth measures the contribution of $i$ to a network-wide smoothness energy.
 
\subsubsection{Neighborhood Residual}
 
Neighborhood residual is the most direct measure of local agreement is the difference between a station's value and the median value of its neighbors,
\begin{equation}
R_i = f_i - \operatorname*{median}_{j\in\mathcal N(i)} f_j ,
\label{eq:resid}
\end{equation}
where $\mathcal N(i)$ is the set of $k$ nearest neighbors of $i$. Use of the median rather than the mean makes $R_i$ resistant to a single anomalous neighbor.
 
\subsubsection{Spatial Gradient}
 
In order to detect abrupt spatial change, we fit a local plane to the neighborhood by distance-weighted least squares,
\begin{equation}
\min_{a,b,c}\;\sum_{j\in\mathcal N(i)} w_{ij}\left[f_j-\left(a + b\,\Delta x_{ij} + c\,\Delta y_{ij}\right)\right]^2,
\qquad w_{ij}=\frac{1}{d_{ij}},
\end{equation}
with $(\Delta x_{ij},\Delta y_{ij})=\mathbf p_j-\mathbf p_i$ expressed in kilometers, and take the gradient magnitude
\begin{equation}
G_i=\lVert\nabla f\rVert_i=\sqrt{\hat b^{\,2}+\hat c^{\,2}}
\end{equation}
in field units per kilometer. A coherent field yields a small and slowly varying $G_i$, whereas a localized disturbance inflates the gradient throughout the surrounding neighborhood rather than at the disturbed station alone.
 
\subsubsection{Prediction Residual}
 
The primary anomaly score predicts a station from all other stations and measures the miss. Using the same inverse-distance-weighted (IDW) interpolator employed for the map products \citep{Shepard1968}, the leave-one-out estimate at station $i$ is
\begin{equation}
\hat f_i=\frac{\sum_{j\neq i} d_{ij}^{-p}\,f_j}{\sum_{j\neq i} d_{ij}^{-p}},\qquad p=2,
\end{equation}
restricted to the eight nearest contributors, and the prediction residual is
\begin{equation}
P_i=f_i-\hat f_i .
\end{equation}
Because $\hat f_i$ never uses $f_i$, a faulty station cannot mask its own fault, and $P_i$ is for that reason the quantity on which the composite index below is built. Kriging interpolation \citep{Cressie1990} can be substituted for IDW without altering the framework.
 
\subsubsection{Graph Smoothness}
 
We quantify the overall spatial smoothness of the observable field using the graph Dirichlet energy \citep{ortega2022introduction}. For this metric, the stations are connected by Gaussian edge weights
\begin{equation}
w_{ij}=\exp\!\left(-\frac{d_{ij}^2}{2\sigma^2}\right),
\end{equation}
in which $\sigma$ is set to the median nearest-neighbor distance of the network. Let $\mathbf{W}$ denote the resulting weight matrix, $\mathbf{D}=\operatorname{diag}(\sum_j w_{ij})$ the degree matrix, and $\mathbf{L}=\mathbf{D}-\mathbf{W}$ the graph Laplacian. The network Dirichlet energy is then
\begin{equation}
J=\mathbf{f}^{\top}\mathbf{L}\mathbf{f}
=\tfrac{1}{2}\sum_i\sum_j w_{ij}(f_i-f_j)^2 ,
\end{equation}
which measures the total weighted variation across neighboring stations. It is small when nearby stations report similar values and grows as local spatial inconsistencies appear. In order to identify where that variation originates, we also compute the contribution of each station,
\begin{equation}
E_i=\sum_j w_{ij}(f_i-f_j)^2 ,
\end{equation}
in which larger values indicate the stations contributing most strongly to the network-wide energy.
 
\subsection{Normalization and the Network Consistency Index}
 
So that metrics and observables can be compared on a common scale, each metric is converted to a $z$-score using the network median and the median absolute deviation (MAD),
\begin{equation}
z(x_i)=\frac{x_i-\operatorname{median}(x)}{1.4826\,\operatorname{MAD}(x)} ,
\end{equation}
so that $|z|\ge3$ marks an outlier under a Gaussian null hypothesis. The composite Network Consistency Index at station $i$ is the weighted quadratic mean of the prediction-residual $z$-scores across the chosen Class A observables $m$,
\begin{equation}
\mathrm{NCI}_i=\sqrt{\frac{\sum_m v_m\,z_{i,m}^2}{\sum_m v_m}},\qquad
z_{i,m}=\bigl|z(P_i^{(m)})\bigr| ,
\label{eq:nci}
\end{equation}
with weights $v_m$ set to unity by default. By construction $\mathrm{NCI}_i\approx1$ for a station that is consistent with its neighborhood across all chosen observables, and $\mathrm{NCI}_i\gg1$ for a station whose observables jointly depart from what the surrounding network predicts.
 
This index is a measure of spatial consistency. It localizes disagreement within the network and acts as an indicator of anomaly. The exact cause can be found through further analysis. This distinction is deliberate, since inconsistency in spatial signature can be produced by a local RFI, a terrestrial spoofer, receiver fault, or an antenna fault.

\section{Dataset}
 
This section describes the dataset, the criterion for including a station in the analysis, data availability, and the processing chain that produces the NCI. The experiment uses two consecutive days of archived observations and broadcast ephemerides.
 
The demonstration draws on observations from the 50-station Alabama DOT (ALDOT) CORS network for two consecutive days, 5 and 6~July~2026, with all epochs referred to Coordinated Universal Time. Each day contains 2{,}880 measurement epochs per station. The data are in RINEX~3.04 format and contain carrier-phase, code, Doppler, and $C/N_0$ measurements at a 30-second interval for GPS, GLONASS, Galileo, and BeiDou \citep{RINEX304}, obtained from the network's post-processing archive. Ionospheric quantities are computed using the thin-shell ionospheric model. While archived data is used for analysis in this study, the CORS network provides a 1 Hz data rate for real-time consumption through the Network Transport of RTCM via Internet Protocol (NTRIP). However, real-time data collection is limited by the agency, which allows only one user to receive one NTRIP stream. Hence, for the purposes of this study, archived data is analyzed, but this method can be translated to real-time operation if all 50 station observations are received simultaneously. 

RINEX observation data for each station are processed into $\Delta$VTEC and ROTI values. From that, the four spatial-consistency metrics $R_i$, $G_i$, $P_i$, and $E_i$ are computed. These metrics are normalized and combined to form the composite index (NCI) in Equation~\eqref{eq:nci}. The result is interpolated to a statewide map, clipped to the state boundary, and masked to the ``coverage-confidence'' region, defined as the area lying within 50\, km of a station, for visualization purposes.

\section{Results}
 
The results are reported in three parts. We first report the data availability and downtime for individual stations. We then apply the four spatial-consistency metrics to two coherent observables to show what each metric contributes and how they differ. Finally, we combine the metrics into the statewide composite index and examine the stations it flags. 

\subsection{Data Availability}
 
An integrity monitor must first establish which stations were in fact delivering data. Forty-nine of the fifty stations delivered $100\%$ of their expected measurement epochs, and the fiftieth, AL92, delivered $99.9\%$, corresponding to a single gap of approximately 2.5 minutes on 5~July. All 50 stations except for the time gap were included in the analysis. High availability is itself the most basic integrity result. The same availability accounting detects genuine outages, and station AL62, which was fully available on both study days, was down on the following day, 7~July. In continuous operation, such a station is excluded by a screening step, as it would otherwise be misread as a spatial anomaly.
 
\subsection{Spatial Consistency Metrics on a Coherent Field}

\begin{figure}[htbp]
    \centering
    \includegraphics[width=0.60\linewidth]{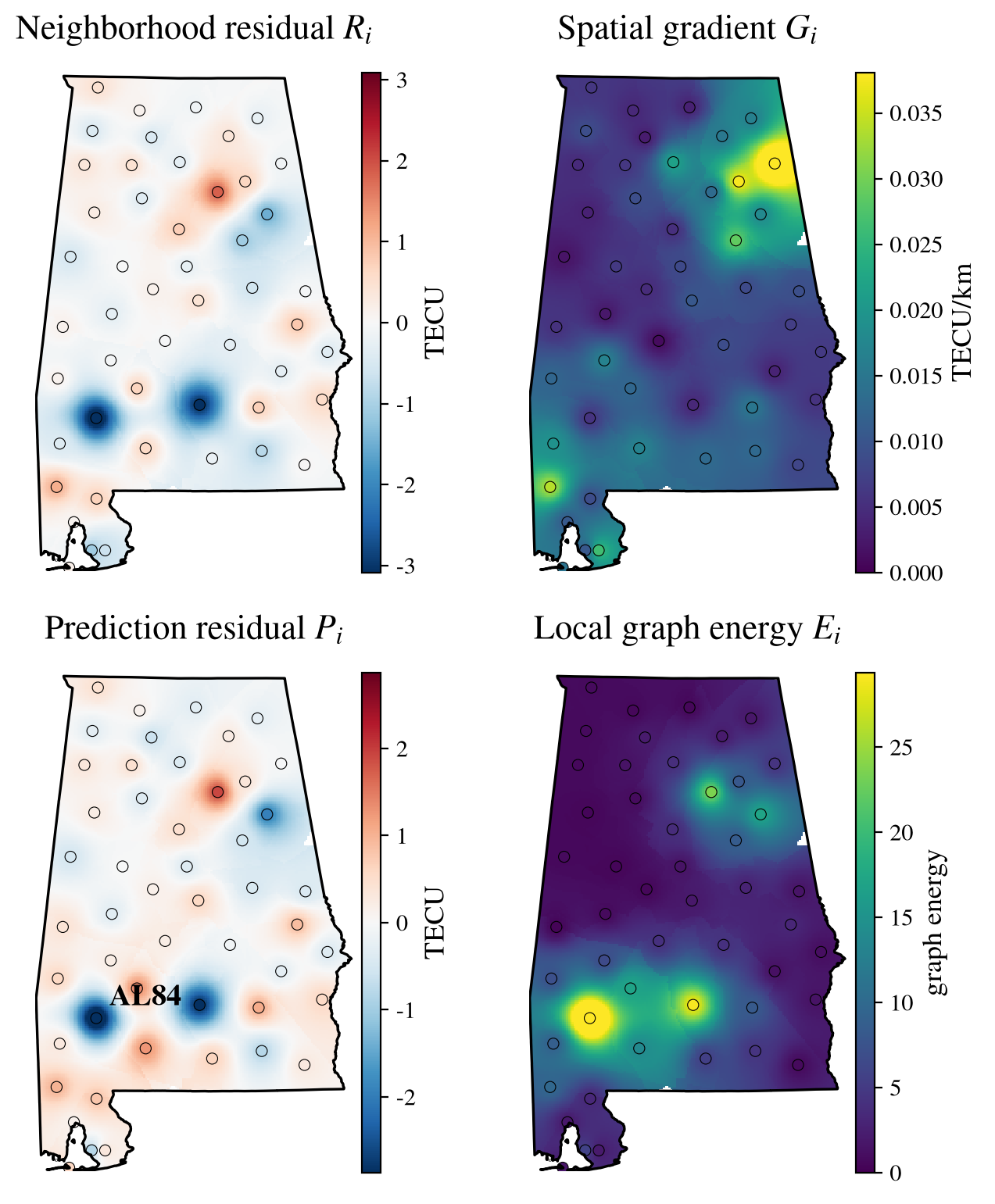}
    \caption{The four NCF spatial-consistency metrics applied to the $\Delta$VTEC field (6~versus 5~July~2026, $N=50$) where AL84 is the strongest outlier.}
    \label{fig:ncf}
\end{figure}

Figure~\ref{fig:ncf} shows the four metrics applied to the day-over-day change in VTEC, for the second study day relative to the first. Across the state $\Delta$VTEC changed by a smooth $+1.2$~TECU at the median, spanning $-2$ to $+3$~TECU, with small gradients and a low total graph Dirichlet energy of $J=139$. This is the nominal behavior that the framework expects of a Class A observable. However, it is worth noting that this method can be applied on much shorter time intervals.

Station AL84, in west-central Alabama, is the clear exception. Its $\Delta$VTEC of $-2.1$~TECU sits against a neighborhood near $+1$~TECU, and this isolated deviation produces the largest neighborhood residual ($|z|=5.8$), the largest prediction residual ($P_i=-3.6$~TECU, $|z|=5.5$), and the highest local graph energy in the network. The gradient panel highlights the ring of neighboring stations, ALGR and ALAS, across which the field changes abruptly, and these two stations carry the next-largest residuals. The four metrics thus give a consistent picture from complementary viewpoints; the residual metrics identify the station, and the gradient metric identifies the neighborhood that the station disturbs.

Further analysis of the AL84 indicates that the anomaly is local to the receiver. It is spatially isolated and not shared by any neighbor; the station shows jumps on $C/N_0$, which is a Class B observable computed outside the spatial test; and on the following day the same station showed a much larger offset of approximately 14~TECU that was uniform across all satellites, which is a DCB signature. What the framework isolates is therefore a genuine anomaly. However, a similar signature could be intentional interference/spoofing. Determining the exact cause of the anomaly requires further investigation using spectrum-monitoring techniques or other signal-level GNSS interference and spoofing detection methods. Nevertheless, the framework presented in this paper identified Station AL84 solely based on its spatial inconsistency with neighboring stations, using only the observations already collected by the statewide CORS network.

\subsection{Statewide Network Consistency Index}

\begin{figure}[htbp]
    \centering
    \includegraphics[width=0.60\linewidth]{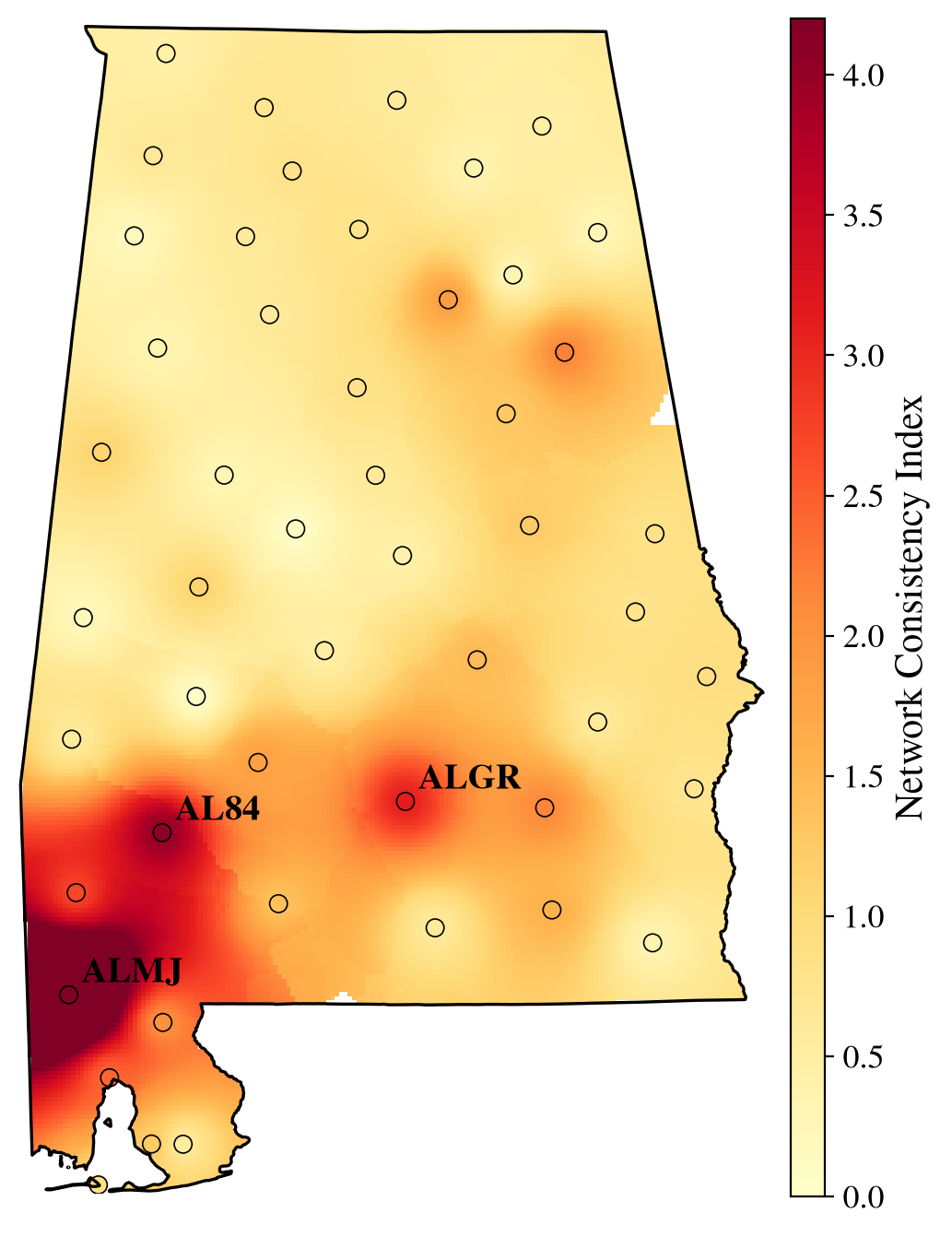}
    \caption{Statewide Network Consistency Index (6~versus 5~July~2026) over the two Class A observables, $\Delta$VTEC and ROTI.}
    \label{fig:nci}
\end{figure}

Figure~\ref{fig:nci} shows the composite index of Equation~\eqref{eq:nci}, computed over the full second study day from the two class~A observables. The $\Delta$VTEC observable was dissected in the previous subsection; the second observable is ROTI, which under the geomagnetically quiet conditions of the study is uniform across the state at a median of $0.09$~TECU/min. The prediction residuals behave similarly to $\Delta$VTEC, with a single outlier at station ALMJ. Combining the two observables, the calculated statewide NCI has a median index of $0.76$ and a ninetieth percentile of $2.2$. Two stations rise well above that background and are labeled on the map, namely ALMJ with an index of $10.6$ and AL84 with an index of $4.1$. Table~\ref{tab:nci} lists the eight least spatially consistent stations together with the observable that dominates each score. The final column reports only evidence computed outside the spatial framework; entries marked ``none available'' indicate a flag that the present dataset cannot confirm or dismiss.

\begin{table}[htbp]
    \caption{The eight least spatially consistent ALDOT stations over the two study days (5--6 July 2026).}
    \label{tab:nci}
    \centering
    \small
    \begin{tabular}{llll}
        \toprule
        Station & NCI & Dominant contributor & External evidence \\
        \midrule
        ALMJ & 10.6 & ROTI ($|z|=15$)              & ROTI 0.29 TECU/min, $3.2$ times network median \\
        AL84 & 4.1  & $\Delta$VTEC ($P_i=-3.6$ TECU) & $C/N_0$ jump; 14 TECU all-satellite bias jump on 7 July \\
        ALGR & 3.1  & $\Delta$VTEC ($|z|=4.4$)      & Not independent; neighbor of AL84 \\
        ALCH & 2.7  & ROTI ($|z|=3.8$)              & None available \\
        AL90 & 2.4  & ROTI ($|z|=3.4$)              & None available \\
        ALAS & 2.2  & $\Delta$VTEC ($|z|=3.1$)      & Not independent; neighbor of AL84 \\
        ALTU & 2.2  & ROTI ($|z|=2.8$)              & None available \\
        AL92 & 2.0  & ROTI ($|z|=2.7$)              & None available \\
        \bottomrule
    \end{tabular}
\end{table}

Of the two, AL84 is the substantive detection. It is driven by $\Delta$VTEC, as dissected above, and it can be explained by evidence external to the spatial test: a simultaneous $C/N_0$ jump and, on the following day, a $\approx14$~TECU offset uniform across all satellites, which is a differential code-bias signature. It is the strong case with multiple indicators.

ALMJ is driven almost entirely by ROTI, at $0.29$~TECU/min against a network median of $0.09$~TECU/min, with a prediction-residual $|z|$ of $15$. The elevation in ROTI value is confined to that single station: its neighbors in every direction, out to tens of kilometers, remain quiet, so this is a genuinely local anomaly rather than a regional ionospheric disturbance, which would be shared, and the framework flags it correctly. Its cause, however, is unresolved. No evidence external to the framework is available for it, and it sits in a sparse part of the network, its nearest neighbor roughly $42$~km away against a network median spacing near $21$~km, so its score carries wider prediction uncertainty than an interior station's. We therefore report ALMJ as a lower-confidence detection. It is a real, spatially isolated anomaly whose cause could be a receiver or antenna fault, local unintentional or intentional RFI, or a spoofing attack. Finding the specific cause would require receiver-level or spectrum-monitoring follow-up. The presented framework successfully isolates it and leaves it open for further investigation.

The remaining six stations are reported for completeness. ALGR and ALAS are driven by $\Delta$VTEC and are the neighbors of AL84 identified by the gradient metric, so their scores are not independent of the AL84 event. ALCH, AL90, ALTU, and AL92 are driven by ROTI at $|z|$ between 2.7 and 3.8, and for these four stations no evidence external to the framework is presently available. 

\section{Conclusions}

This study demonstrates that statewide CORS networks, traditionally deployed for high-accuracy GNSS positioning services, can be extended into regional integrity monitoring infrastructures. This study demonstrates that the same infrastructure can serve as a foundation for a regional GNSS integrity monitoring system. In this study, we introduced the NCF, which evaluates each station's agreement with its spatial neighborhood using four complementary graph-based consistency metrics and aggregates these metrics into a Network Consistency Index (NCI) that quantifies station-level spatial consistency. Applied to the Alabama DOT network, the framework isolates two localized anomalies. These results show that deviations from regional spatial consistency can be detected using only observations already collected by statewide CORS networks.

The framework requires no additional hardware and can operate continuously using existing network observations, extending current network monitoring beyond positioning performance to GNSS integrity. Rather than relying solely on receiver-level diagnostics or historical behavior at individual stations, the proposed approach exploits the spatial coherence naturally present in regional GNSS observables to identify stations that disagree with their surrounding network. This provides transportation agencies with a practical mechanism for prioritizing investigation of localized interference, potential spoofing events, and receiver-related faults. By leveraging spatial redundancy across distributed CORS infrastructure, the approach presented in this paper, provides an additional layer of cyber physical resilience for transportation systems relying on GNSS services.

Several limitations define the scope of the present work. The demonstration focuses on nominal operation and naturally occurring anomalies rather than controlled spoofing experiments. Only two days of observations were analyzed, so seasonal, geomagnetic, and long-term variability remain to be evaluated. Detection performance also depends on network geometry, with reduced confidence near sparsely instrumented boundaries, suggesting benefits from adaptive neighborhood weighting and cross-state data sharing. Future research will evaluate longer observation periods, controlled spoofing experiments using a virtual reference station, and real-time implementation within operational statewide CORS networks.

\FloatBarrier  

\section*{FUNDING}
This work is based upon the work supported by the National Center for Transportation Cybersecurity and Resiliency (TraCR) (a U.S. Department of Transportation National University Transportation Center) headquartered at Clemson University, Clemson, South Carolina, USA, and the National Science Foundation (NSF) (Award \# 2340456). 

\section{Acknowledgments}
Any opinions, findings, conclusions, and recommendations expressed in this material are those of the author(s) and do not necessarily reflect the views of funding agencies, and the U.S. Government assumes no liability for the contents or use thereof.

We have used generative AI (Claude Opus 4.8) for editorial purposes.

\section*{AUTHOR CONTRIBUTIONS}
The authors confirm their contributions to the paper as follows:
study conception and design: Minhaj Uddin Ahmad, Sagar Dasgupta, Mizanur Rahman;
methodology, formal analysis, software, and visualization: Minhaj Uddin Ahmad;
validation: Minhaj Uddin Ahmad, Muhammad Sami Irfan, Sagar Dasgupta, Mizanur Rahman;
writing -- original draft: Minhaj Uddin Ahmad;
writing -- review and editing: Sagar Dasgupta, Muhammad Sami Irfan, Mizanur Rahman, Mashrur Chowdhury;
All authors reviewed the results and approved the final version of the manuscript.

\section*{DECLARATION OF CONFLICTING INTERESTS}
The authors declared no potential conflicts of interest with respect to the research, authorship, and/or publication of this article.

\newpage
\bibliographystyle{trb}
\bibliography{main}
\end{document}